\documentclass[aps,pre,floats,floatfix,twocolumn]{revtex4-1}

\RequirePackage{lineno}

\usepackage{graphicx}
\usepackage{epstopdf}
\graphicspath{{./Figs/}{./figs/}}

\usepackage{latexsym}
\usepackage{amsmath}
\usepackage{amssymb}
\usepackage{bm}
\usepackage{wasysym}

\usepackage{ifthen}
\usepackage{color}
\usepackage[colorlinks,allcolors=blue,bookmarks=true]{hyperref}

\newcommand{\trc}{\mbox{trace}}

\newcommand{\amatrix}[1]{\begin{matrix} #1 \end{matrix}} 
\newcommand{\eexp}[1]{\mathrm{e}^{#1}}
\newcommand{\pd}[2]{\frac{\partial #1}{\partial #2}}

\newcommand{\braket}[1]{ \left\langle #1 \right\rangle}

\newcommand{\be}[1]{\begin{eqnarray}{\label{e#1}}} 
\newcommand{\beq}{\begin{eqnarray}}
\newcommand{\eeq}{\end{eqnarray}} 

\newcommand{\hide}[1]{}
\newcommand{\Eq}[1]{\textcolor{blue}{{Eq.}\!\!~(\ref{#1})}} 

\newcommand{\Fig}[1]{\textcolor{blue}{Fig.}\!\!~\ref{#1}}
\definecolor{myred}{rgb}  {0.5,0.0,0.0}

\newcommand{\hrefl}[2]{\href{#2}{(#1)}}

\newcommand{\tr}{\text{Tr}\,}

\begin{document}

\title{The lognormal-like statistics of a stochastic squeeze process}

\author{Dekel Shapira, Doron Cohen}

\affiliation{Department of Physics, Ben-Gurion University of the Negev, Beer-Sheva 84105, Israel}

\begin{abstract}
We analyze the full statistics of a stochastic squeeze process.
The model's two parameters 
are the bare stretching rate~$w$, 
and the angular diffusion coefficient~$D$.    
We carry out an exact analysis to determine 
the drift and the diffusion coefficient of $\log(r)$, 
where $r$ is the radial coordinate. 
The results go beyond the heuristic lognormal description
that is implied by the central limit theorem. 
Contrary to the common ``Quantum Zeno'' approximation, 
the radial diffusion is not simply $D_r = (1/8)w^2/D$, 
but has a non-monotonic dependence on $w/D$.
Furthermore, the calculation of the radial moments 
is dominated by the far non-Gaussian tails 
of the $\log(r)$ distribution.   
\end{abstract}

\maketitle

\section{Introduction}

In this paper we analyze the full statistics of a 
physically-motivated  stochastic squeeze process 
that is described by the Langevin (Stratonovich) equation 
\beq \nonumber
\dot{x}  \ &=& \ \ \ w x  \ - \ \omega(t) y  
\\ \label{e1}
\dot{y}  \ &=& \ -w y  \ + \ \omega(t) x 
\eeq
where the rotation frequency $\omega(t)$ is a zero mean
white noise with fluctuations:
\be{2}
\braket{\omega(t') \omega(t'')} \ = \ 2D \delta(t'-t'')
\eeq
Accordingly the model has two parameters: 
the angular diffusion coefficient~$D$ of the polar {\em phase}, 
and the bare stretching rate~$w$ of the radial 
coordinate ${r=\sqrt{x^2+y^2}}$.
In a physical context the noise arises due to the 
interaction with environmental degrees of freedom, 
typically modeled as an harmonic bath of ``phonons". 
Hence we can assume for it a Gaussian-like distribution 
with bounded moments. The white noise assumption 
means that the correlation time is very short, hence 
the Stratonovich interpretation of \Eq{e1} is in order,
as argued, for example, by Van Kampen \cite{VanKampen1981}.

The squeeze operation is of interest in many fields 
of science and engineering, but our main motivation originates from the 
quantum mechanical arena, where it is known as parametric amplification.
In particular it describes the dynamics of a Bosonic Josephson Junction (BJJ)  
given that all the particles are initially condensed in the upper orbital. 
Such preparation is {\em unstable} \cite{Chuchem10,chapman},  
but it can be stabilized by introducing frequent {\em measurements} or by introducing {\em noise}. 
This is the  so-called ``quantum Zeno effect'' (QZE) \cite{QZE1,QZE3,QZE4,Kofman2,Gordon1}.
The manifestation of the QZE in the BJJ context  
has been first considered in \cite{QZESQ1,QZESQ2},
and later in \cite{KhripkovVardiCohen2012}.

The main idea of the QZE is usually explained as follows:  
The very short-time decay of an initial preparation 
due to a constant perturbation is described 
by the survival probability ${\mathcal{P}(t)=1-(vt)^2}$, 
where~$v$ is determined by pertinent couplings to the other eigenstates;  
Dividing the evolution into $\tau$-steps, and assuming 
a projective measurement at the end of each step one obtains
\beq \nonumber
\mathcal{P}(t) &\approx& \left[\mathcal{P}(\tau)\right]^{t/\tau}
\approx \left[1-(v\tau)^2\right]^{t/\tau}  
\approx \exp\left[-(v^2\tau)t\right]
\eeq 
The common phrasing is that frequent measurements (small $\tau$) 
slow down the decay process due to repeated ``collapse" of the wavefunction.    
Optionally one considers a system that is coupled to the environment. 
Such interaction is formally similar to a continuous measurement process, 
that is characterized by a dephasing time~$\tau$. 
In the latter case the phrasing is that the introduction 
of ``noise" leads to the slow-down of the decay process. 
Contrary to simple minded intuition, stronger noise leads 
to slower decay.

At this point one might get the impression that the QZE 
is a novel ``quantum" effect, that has to do with mysterious 
collapses, and that such effect is not expected to arise 
in a ``classical" reality.  
Such conclusion is in fact wrong: whenever the the system of 
interest has a meaningful classical limit, the same Zeno effect 
arises also in the classical analysis.   
This point has been emphasized by Ref.\cite{KhripkovVardiCohen2012}
in the context of the BJJ.
It has been realized that the QZE is the outcome 
of the classical dynamics that is is generated by \Eq{e1}, 
where the ${(x,y)}$ are local canonical conjugate coordinates 
in the vicinity of an hyperbolic (unstable) fixed-point in phase space. 
The essence of the QZE in this context  
is the observation that the introduction of the noise 
via the phase-variable leads to slow-down 
of the radial spreading. 
For strong noise (large~$D$ in \Eq{e2}), 
the radial spreading due to~$w$ is inhibited. 
Using quantum terminology this translates 
to suppression of the decoherence process.

From pedagogical point of view it is useful to note that 
the dynamics of the BJJ is formally similar to that of a mathematical pendulum.
Condensation of all the particle in the upper orbital 
is formally the same as preparing the pendulum in the upper position. 
Such preparation is unstable. 
If we want to stabilize the pendulum in the upper position  
we have the following options: 
(i)~Introducing periodic driving that leads to the Kapitza effect; 
(ii)~Introducing noisy driving that leads to a Zeno effect.
We note that the Kapitza effect in the BJJ context 
has been discussed in \cite{kapitza}, 
while our interest here is in the semiclassical perspective 
of the QZE that has been illuminated in \cite{KhripkovVardiCohen2012}.

Experiments with cold atoms are state of the art \cite{BHH1,BHH2}.
In such experiments it is common to perform a ``fringe visibility''
measurement, which indicates the condensate occupation. 
The latter is commonly quantified in terms of a function ${\mathcal{F}(t)}$.  
For the initial coherent preparation ${\mathcal{F}(t)=1}$,   
while later (ignoring quantum recurrences) it decays to a smaller value.  
Disregarding technical details the {\em standard} QZE argument 
implies an exponential decay 
\beq
\mathcal{F}(t) \ \ = \ \ \exp\left\{-\frac{1}{N} \mathcal{S}(t)\right\}
\eeq
where $N$ is the number of condensed bosons, and 
\be{3}
\mathcal{S}(t) \ \ = \ \ \left(\frac{w^2}{D}\right) t 
\eeq 
The key realizations of Ref.\cite{KhripkovVardiCohen2012} is 
that  $\mathcal{S}(t)$ is in fact the radial spreading in a stochastic process that is described by \Eq{e1}.

A practical question arises, whether 
the heuristic QZE expression for $\mathcal{S}(t)$
is {\em useful} in order to describe the actual 
decay of the one-body coherence.
The answer of Ref.\cite{KhripkovVardiCohen2012} was:
(i)~The heuristic result is correct only 
for a very strong noise (small $w/D$), and holds 
only during a very short time.
(ii)~Irrespective of correctness, it is unlikely 
to obtain a valid estimate for $\mathcal{S}(t)$  
in a realistic measurement, because the statistics
is log-normal, dominated by far tails.

On the quantitative side, Ref.\cite{KhripkovVardiCohen2012} 
was unable to provide an analytical theory for 
the lognormal statistics of the spreading.    
Rather it has been argued that the $\ln(r)$~distribution 
has some average ${\mu \propto t}$, 
and some variance ${\sigma^2 \propto t}$.
The radial stretching rate $w_r$    
and a radial diffusion coefficient $D_r$
were determined {\em numerically} from the assumed time dependence: 
\beq
\mu \ &=& \ w_r t \\
\sigma^2 \ &=& \ 2D_r t 
\label{eq:sigma-sqr-2dt}
\eeq
From the lognormal assumption it follows that
\be{LN}
\mathcal{S}(t) \ = \  e^{4D_r t + 2w_r t} - 1  
\eeq
For strong noise the following asymptotic results have been obtained: 
\be{9}
w_r  \ &\sim& \ \frac{w^2}{4D} 
\\ \label{eq:D-r-large-d}
D_r  \ &\sim&  \ \frac{w^2}{8D} 
\eeq  
These approximations are satisfactory for ${w/D \ll 1}$,   
but fail miserably otherwise. 
We also see that \Eq{eLN} reduces to \Eq{e3}
in this strong noise limit, for a limited duration of time.
Note that \Eq{eLN} is not identical with the expression
that has been advertised in \cite{KhripkovVardiCohen2012}
for reasons that will be discussed in the concluding section.

{\bf Outline.-- } 
The QZE motivation for the analysis of \Eq{e1} is introduced in Sections Sections~II.
Numerical results for the radial spreading due to such process are presented in Section~III.  
Our objective is to find explicit expression for $w_r$ and $D_r$, 
and also to characterize the full statistics of $r(t)$  
in terms of the bare model parameters ${(w,D)}$.  
The first step is to analyze the phase randomization in Sections~IV, 
and to discuss the implication of its non-isotropic distribution in Section~V.
Consequently the exact calculation of the $\ln(r)$~diffusion 
is presented in Sections~VI and~VII.
In Sections~VIII we clarify that the statistics of $r(t)$ is in fact 
a {\em bounded} lognormal distribution. It follows that the $r$~moments 
of the spreading, unlike the $\ln(r)$~moments, 
cannot be deduced directly from our results for $w_r$ and $D_r$.  
Nevertheless, in Section~IX we find the $r$~moments 
using the equation of motion for the moments. 
Finally in Section~X we come back to the discussion 
of the QZE context of our results. On the one hand 
we note that \Eq{eLN} should be replaced 
by a better version that takes into account 
the deviations from the lognormal statistics.
But the formal result for $\mathcal{S}(t)$ has no experimental significance: 
the feasibility of experimental $\mathcal{S}(t)$ determination is questionable, 
because averages are sensitive to the far tails. 
Rather, in a realistic experiment it is feasible to accumulate statistics 
and to deduce what are $w_r$ and $D_r$, 
which can tested against our predictions.
Some extra details regarding the QZE perspective and
other technicalities are provided in the Appendices.

\section{Semiclassical perspective}

In the present section we clarify the semiclassical perspective for 
the QZE model, and motivate the detailed analysis of \Eq{e1}.
The subsequent sections are written in a way that is independent 
of a specific physical context. We shall come back to the discussion 
of the QZE in the concluding section, where the implications of 
our results are summarized.

For a particular realization of $\omega(t)$ the evolution that is 
generated by \Eq{e1} is represented by a symplectic matrix
\beq
\left(\amatrix{x(t) \cr y(t)}\right) \ \ = \ \ \bm{U} \left(\amatrix{x_0 \cr y_0}\right)
\eeq
The matrix is characterized by its trace ${a=\trc(\bm{U})}$.
If ${|a|< 2}$ it means elliptic matrix (rotation). If $|a|> 2$ 
it means hyperbolic matrix. In the latter case, 
the radial coordinate~$r$ is stretched
in one major direction by some factor $\exp(\alpha)$, 
while in the other major direction 
it is squeezed by factor $\exp(-\alpha)$. 
Hence ${a=\pm 2\cosh(\alpha)}$. 
If we operate with $\bm{U}$ on an initial isotropic cloud 
that has radius $r_0$, then we get a stretched cloud 
with ${\braket{r^2}=\mathcal{A} \, r_0^2}$, 
where ${\mathcal{A}=\cosh(2\alpha)}$.       
For more details see Appendix~A.
The numerical procedure of generating a stochastic process
that is described by \Eq{e1} is explained in Appendix~B.  
Rarely the result is a rotation. So from now on we refer to it as ``squeeze".

The initial preparation can be formally described as a minimal wavepacket 
at the origin of phase-space. The local canonical coordinates are ${(x,y)}$, 
or optionally one can use the polar coordinates ${(\varphi,r)}$. 
The initial spread of the wavepacket is  ${\braket{r^2} = \hbar}$.
In the case of a BJJ the dimensionless Planck constant 
is related to the number of particles, namely ${\hbar=2/N}$.   
In the absence of noise (${D=0}$) the wavepacket is stretched   
exponentially in the~$x$ direction, which implies a very fast decay 
of the initial preparation. This decay can be described by 
functions $\mathcal{P}(t)$ and $\mathcal{F}(t)$ 
that give the survival probability of the initial state,  
and the one-body coherence of the evolving state.  
For precise definitions see Appendix~C. 
Note that $\mathcal{F}(t)$ is defined as the length 
of the Bloch vector, normalized such that ${\mathcal{F}(t)=1}$
for the initial coherent state.

We now consider the implication of having a noisy dephasing term (${D>0}$).
The common perspective is to say that this noise 
acts like a measurement of the $r$ coordinate, 
which randomizes the phase $\varphi$ over a time scale ${\tau\sim 1/D}$, 
hence introducing a ``collapse'' of the wave-function. 
The succession of such interventions (see Appendix~C)
leads to a relatively slow exponential decay of the coherence, 
namely ${ \mathcal{F}(t) = \exp\left\{-(\hbar/2)\mathcal{S}(t)\right\}}$, 
where $\mathcal{S}(t)$ is given by \Eq{e3}.
The stronger the noise ($D$), the slower is the decay 
of~$\mathcal{F}(t)$. Similar observation applies to~$\mathcal{P}(t)$.   
Using a semiclassical perspective \cite{KhripkovVardiCohen2012} 
it has been realized that 
\beq
\mathcal{S}(t) \ \ = \ \ \mathcal{A}(t)-\mathcal{A}(0)
\eeq
Note that by definition $\hbar\mathcal{A}(t)$ is the spread ${\braket{r^2}}$
of the evolving phase-space distribution, where $\mathcal{A}(t)$ is 
normalized such that ${\mathcal{A}(0)=1}$.

The well known QZE expression \Eq{e3}, in spite of its popularity, 
poorly describes the decoherence process~\cite{KhripkovVardiCohen2012}. 
In fact, it agrees with numerical simulations 
only for extremely short times for which ${(w^2/D) t \ll 1}$.    
The semiclassical explanation is as follows: 
In each $\tau$-step of the evolution the phase-space distribution 
is stretched by a random factor ${\lambda_n = \exp[\alpha_n]}$, 
where the $\alpha_n$ are uncorrelated random variables. 
Hence by the central limit theorem 
the product ${\lambda=\lambda_t...\lambda_2\lambda_1}$ 
has lognormal distribution, 
where $\log(\lambda)$ has some average~${\mu \propto t}$ 
and variance~${\sigma^2 \propto t}$ that determine an $\mathcal{A}(t)$ 
and hence $\mathcal{S}(t)$
that differs from the naive expression of \Eq{e3}.     
The essence of the QZE is that $\mu$ and $\sigma^2$ are inversely proportional 
to the intensity of the erratic driving. 
Consequently one has to distinguish between 3 time scales: 
the ``classical" time for phase ergodization ${\tau \sim D^{-1}}$ 
which is related to the angular diffusion;  
the ``classical" time for loss of isotropy ${t_r \sim (w^2/D)^{-1}}$
that characterizes the radial spreading;   
and the ``quantum" coherence time~${t_c\sim (1/\hbar) t_r}$, 
after which ${\mathcal{F}(t) \ll 1}$.

In \cite{KhripkovVardiCohen2012} the time dependence of $\mu$ and $\sigma$ has been determined numerically.  
Here we would like to work out a proper analytical theory.
It turns out that a quantitative analysis of the stochastic 
squeezing process requires to go beyond the above heuristic description. 
The complication arises because what we have is not multiplication of random number, 
but multiplication of random matrices. 
Furthermore we shall see that the calculation of moments  
requires to go beyond central limit theorem, because they 
are dominated by the far tails of the distribution.

In the concluding section~X we shall clarify that from an experimental point 
of view the formal expression ${\mathcal{F}(t) = \exp\left\{-(\hbar/2)\mathcal{S}(t)\right\}}$ is not very useful. 
For practical purpose it is better to consider the {\em full} statistics of the Bloch-vector, 
and to determine~$\mu$ and~$\sigma$ via a standard fitting procedure.

\section{Preliminary considerations}

\begin{figure}[b]
\includegraphics[width=7cm]{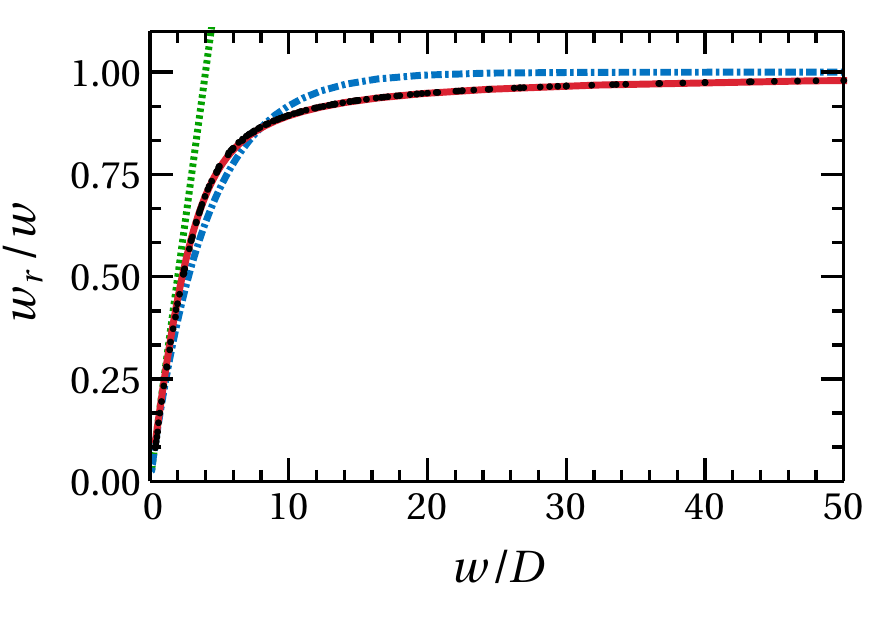}

\caption{ 
Scaled stretching rate $w_r/w$ versus $w/D$.
The numerical results (black symbols)
are based on simulations with 2000 realizations.
The lines are for the naive result \Eq{e9} (green dotted); 
the exact result  \Eq{e16} (red solid); 
and its practical approximation  \Eq{eq:wr-exact-interpolation} (blue dashed-dotted). 
For large values of $w/D$ we get $w_r/w = 1$, as for a pure stretch.
\hfill
} 

\label{fig:wr-vs-d} 

\ \\

\includegraphics[width=7cm]{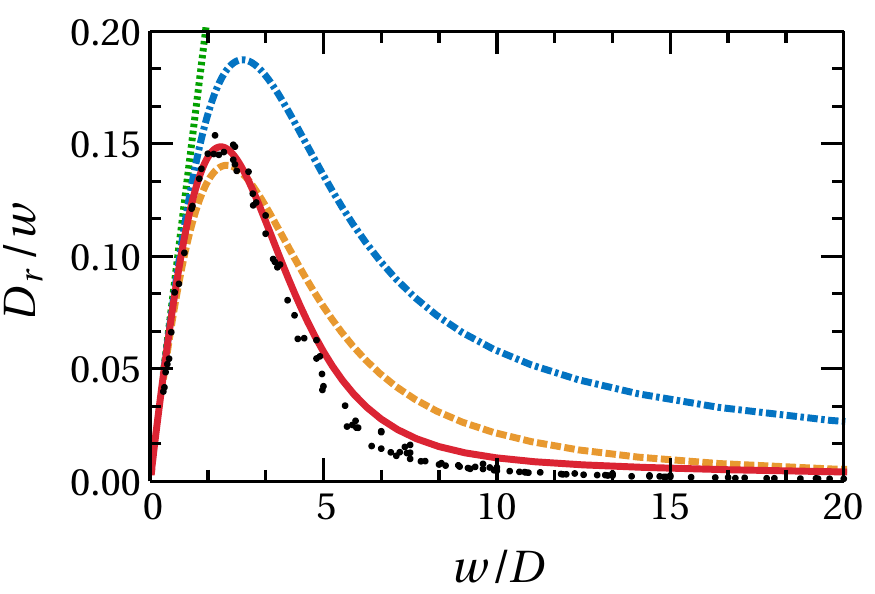}

\caption{ 
Scaled diffusion coefficient $D_r/w$ versus $w/D$. 
The numerical results (black symbols)
are based on simulations with 2000 realizations.
The lines are for the naive result \Eq{eq:D-r-large-d} (green dotted); 
the exact result  \Eq{eq:Dr-exact-result} (red solid); 
and the approximation \Eq{eq:dr-large-d-approximation} 
with ${\tau=1/(2D)}$ (blue dashed-dotted), 
and with \Eq{eq:tau-const-from-exact-result} (dashed orange line).
\hfill
} 

\label{fig:dr-vs-w/d} 
\end{figure}

Below we are not using a matrix language, but address directly 
the statistical properties of an evolving distribution.   
In ${(\varphi,r)}$ polar coordinates \Eq{e1} takes the form 
\beq
\dot{\varphi} & \ = \ &  - w\sin(2\varphi)  \ + \  \omega(t)  \label{eq:langevin-phi}\\
\dot{r} & \ = \ &  [w\cos(2 \varphi)] \ r   \label{eq:langevin-r}
\eeq
We see the equation for the phase decouples, 
while for the radius
\beq
\label{eq:langevin-ln-r}
\frac{d}{dt}\ln(r(t)) \ \ = \ \ w\cos(2 \varphi)
\eeq 
The RHS has some finite correlation time ${\tau \sim 1/D}$, 
and therefore $\ln(r)$ is like a sum of $t/\tau$ uncorrelated 
random variables. It follows from the central limit theorem 
that for long time the main body of the $\ln(r)$ distribution 
can be approximated by a {\em normal} distribution, 
with some average ${\mu \propto t}$, 
and some variance ${\sigma^2 \propto t}$.
Consequently we can define a radial stretching rate $w_r$    
and a radial diffusion coefficient $D_r$ via \Eq{eq:sigma-sqr-2dt}.

Our objective is to find explicit expression for  $w_r$ and $D_r$, 
and also to characterize the full statistics of $r(t)$  
in terms of the bare model parameters ${(w,D)}$.  
We shall see that the statistics of $r(t)$ is described 
by a {\em bounded lognormal distribution}.     

Some rough estimates are in order. 
For large $D$ one naively assumes 
that due to ergodization of the phase ${\mu = \braket{\cos(2 \varphi)}w }$ is zero,     
while ${\sigma^2 \sim (w \tau)^2 (t/\tau)}$. 
Hence one deduces that ${w_r \rightarrow 0}$ while ${D_r \propto w^2/D}$. 
A more careful approach \cite{KhripkovVardiCohen2012}  
that takes into account the non-isotropic distribution 
of the phase gives the asymptotic results \Eq{e9} and \Eq{eq:D-r-large-d}.
The dimensionless parameter that controls the accuracy 
of this result is $w/D$. These approximations are satisfactory 
for ${w/D \ll 1}$, and fails otherwise, see \Fig{fig:wr-vs-d} and \Fig{fig:dr-vs-w/d}.
For large $w/D$ we get $w_r \rightarrow  w$,  while $D_r \rightarrow 0$.

\section{Phase ergodization}

\begin{figure}[t]
\includegraphics[width=7cm]{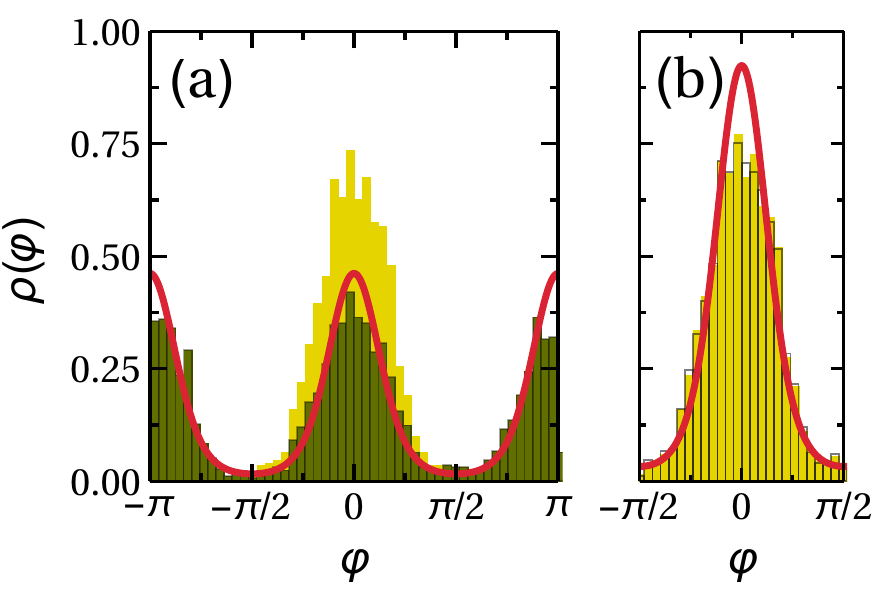}

\caption{
{ \bf (a) } Phase distribution for $(w/D) = 10/3$ after time $(wt) = 6$, with initial conditions $\varphi = 0$ (filled, yellow) and $\varphi = \pi/2$ (green bars) with 2000 realizations. For larger times, both reach the steady state of \Eq{eq:ro-fi-steady-state} (red line). { \bf(b)} The distributions of $\varphi$ modulo $\pi$. 
\hfill
} 

\label{fig:phase-distribution} 
\end{figure}

The Fokker-Planck equation (FPE) that is associated with \Eq{eq:langevin-phi} is  
\beq
\label{eq:fokker-planck-phi}
\frac{\partial \rho}{\partial t} \ = \ \frac{\partial}{\partial \varphi} 
\left[ \left(D \frac{\partial}{\partial \varphi} + w\sin(2\varphi) \right) \rho \right]
\eeq
It has the canonical steady state solution
\beq
\rho_{\infty}(\varphi)  \ \ \propto \ \ \exp\left[ \frac{w}{2D} \cos(2\varphi) \right]  
\label{eq:ro-fi-steady-state}
\eeq
If we neglect the cosine potential in \Eq{eq:fokker-planck-phi} then the time for ergodization 
is ${\tau_{\text{erg}} \sim 1/D}$. But if $w/D$ is large we have to 
incorporate an activation factor, accordingly 
\beq
\tau_{\text{erg}} \ \ = \ \ \frac{1}{D} \ \exp\left[ \frac{w}{D} \right] 
\eeq    
\Fig{fig:phase-distribution}(a) shows the distribution of the phase for two different initial conditions, as obtained by a finite time numerical simulation.  It is compared with the steady state solution. The dynamics of $r$ depends only on $2 \varphi$, and is dominated by the distribution at the vicinity of $\cos(2 \varphi) \sim 1$. We therefore display in \Fig{fig:phase-distribution}(b) the distribution of $\varphi$ modulo $\pi$.  We deduce that the transient time of the $\ln(r)$ spreading is much shorter than $\tau_{\text{erg}}$.

\begin{figure}[t]
\includegraphics[width=7cm]{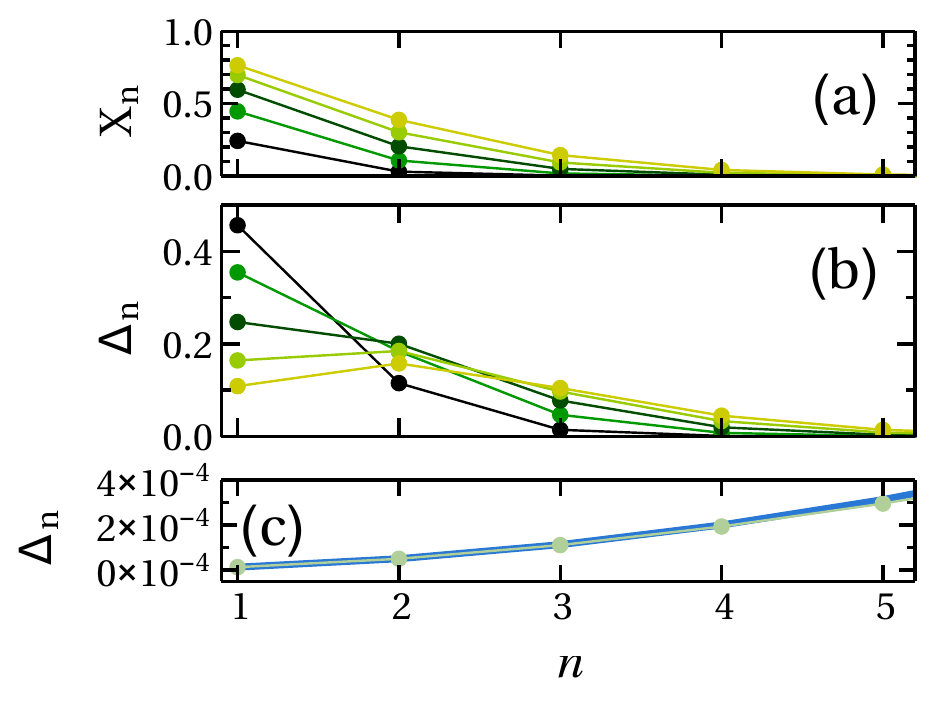}

\caption{
{\bf (a)} 
The values of $X_n$ versus $n$ for some values of $w/D$. 
From bottom to top ${w/D = 1,2,3,4,5}$.
{\bf (b)}
The values of $\Delta_n$ versus $n$ for the same values of $w/D$,  
from top to bottom at $n{=}1$. 
{\bf (c)}
$\Delta_n$ versus $n$ for large $w/D$.  Here $w/D = 400$.
The asymptotic approximation \Eq{e19} is indicated by blue line.
\hfill
} 

\label{fig:deltan-xn-various-n} 
\end{figure}

For the later calculation of $w_r$ we have to know 
the moments of the angular distribution. 
From \Eq{eq:ro-fi-steady-state} we obtain: 
\beq
\label{eq:X-n-result}
X_n \ \ \equiv \ \ 
\braket{\cos(2n\varphi)}_{\infty} \ \ = \ \ \frac{I_n\left(\frac{w}{2D}\right)}{I_0\left(\frac{w}{2D}\right)} 
\eeq
Here $I_n(z)$ are the modified Bessel functions.
For small $z$ we have ${I_n(z) \approx [1/n!](z/2)^n}$, 
while for large $z$ we have ${I_n(z) \approx (2\pi z)^{-1/2} e^z}$.
The dependence of the $X_n$ on $n$ for representative values 
of $w/D$ is illustrated in the upper panel of \Fig{fig:deltan-xn-various-n}.

For the later calculation of $D_r$ we have to know also 
the temporal correlations. We define 
\be{15}
C_n(t) =  
\braket{\cos(2n\varphi_t) \cos(2\varphi)}_{\infty} - X_nX_1
\ \ 
\eeq
where a constant is subtracted such that $C_n(\infty)=0$.
We use the notations 
\beq
\label{eq:cn-integral}
c_n \ \ \equiv \ \ \int_{0}^{\infty} C_n(t) dt
\eeq
and 
\beq
\label{eq:delta-n-def}
\Delta_n \ \equiv \  C_n(0) \ = \ \frac{1}{2}\left(X_{n+1}+X_{n-1}\right) - X_nX_1
\ \ \ \ 
\eeq
In order to find an asymptotic expression we use
\beq\nonumber
I_n(z) \approx \frac{e^{z}}{\sqrt{2 \pi z}}\left[1 - \frac{4 n^2 {-} 1}{(8z)} +   \frac{(4 n^2 {-} 1) (4 n^2 {-} 9)}{2 (8 z)^2} \right]
\eeq
and get  
\be{19}
\Delta_n \ \approx \ 2 \left( \frac{w}{D} \right)^{-2} n^2  
\ \ \ \ \ \ \ \mbox{for} \ \ \left(\frac{w}{D}\right) \gg 1
\eeq
The dependence of the $\Delta_n$ on $n$ for representative values 
of $w/D$ is illustrated in the lower panel of \Fig{fig:deltan-xn-various-n}.

\section{Radial spreading}

If follows from \Eq{eq:langevin-ln-r} that the radial stretching rate is 
\be{16}
w_r \ \ = \ \ w \braket{\cos(2\varphi)}_{\infty} \ \ = \ \ X_1 w 
\eeq
A rough interpolation for $X_1$ that is based on 
the asymptotic expressions for the Bessel functions 
in \Eq{eq:X-n-result} leads to the following approximation 
\beq
\label{eq:wr-exact-interpolation}
w_r \ \ \approx \ \ \ w \left[ 1-\exp\left(-\frac{w}{4D}\right) \right]  
\eeq
The exact result as well as the approximation 
are illustrated in \Fig{fig:wr-vs-d} 
and compared with the results of numerical simulations.

For the second moment it follows from \Eq{eq:langevin-ln-r} 
that the radial diffusion coefficient is 
\beq
\label{eq:Dr-integral}
D_r \ \ = \ \ w^2 \int_{0}^{\infty} C_1(t) dt \ \ = \ \ c_1 w^2
\eeq 
If we assume that the ergodic angular distribution is isotropic, 
the calculation of $C_1(t)$ becomes very simple, namely,  
\beq
\label{eq:C1-naive-result}
C_1(t) \ = \ \frac{1}{2} \braket{\cos{2(\varphi_t-\varphi_0)}} 
\ = \ \frac{1}{2} \eexp{-4 D |t|}
\eeq
This expression implies a correlation time ${\tau=1/(2D)}$, 
such that ${c_1=(1/2)\Delta_1\tau}$ is half the ``area''  of 
the correlation function whose ``height'' is ${\Delta_1=1/2}$.
Thus we get for the radial diffusion coefficient ${D_r = w^2/(8D)}$.

But in fact the ergodic angular distribution is not isotropic, 
meaning that $X_1$ is not zero, and ${\Delta_1<1/2}$.
If $w$ is not too large we may assume that 
the correlation time $\tau$ is not affected. 
Then it follows that a reasonable approximation for 
the correlation function is 
\beq
C_1(t) \ \ \approx  \ \Delta_1 \eexp{-2|t|/\tau} 
\eeq
leading to 
\beq
\label{eq:dr-large-d-approximation}
D_r \ \ \approx \ \ \frac{1}{2}\Delta_1 \tau w^2 \ \ = \ \ \Delta_1 \frac{w^2}{4D}
\eeq
This approximation is compared to the exact result that 
we derive later in \Fig{fig:dr-vs-w/d}.
Unlike the rough approximation ${D_r = w^2/(8D)}$,  
it captures the observed non-monotonic dependence 
of $D_r$ versus $w$, but quantitatively it is an over-estimate.

\section{The exact calculation of the diffusion coefficient}

We now turn to find an exact expression 
for the diffusion coefficient \Eq{eq:Dr-integral}  
by calculating $c_1$ of \Eq{eq:cn-integral}.
Propagating an initial distribution ${\rho_0(\varphi)}$ 
with the FPE \Eq{eq:fokker-planck-phi} we define the moments: 
\beq
\nonumber
x_n \ &=& \ \braket{ \cos(2 n\,\varphi_t) }_0 
\ = \ \braket{ \cos(2 n\,\varphi) }_t 
\\ \label{eq:xn-average}
\ &=& \  \int \cos(2 n\,\varphi)  \ \rho_t(\varphi) d\varphi 
\eeq
The moments equation of motion resulting from the FPE is \cite{risken1984fokker}: 
\beq
\label{eq:d-dt-xn-recursive}
\frac{d}{dt} x_ n \ \ = \ \ -\Lambda_n \, x_n \ + \ W_n  \left( x_{n-1} - x_{n+1} \right) 
\eeq
where ${\Lambda_n=4Dn^2}$ and $W_n=wn$.  
Due to ${\Lambda_0=W_0=0}$ the zeroth moment ${x_0=1}$ does not change in time.
Thus the rank of \Eq{eq:d-dt-xn-recursive} is less than its dimension 
reflecting the existence of a zero mode $x_n=X_n$ 
that corresponds to the steady state of the FPE. 
We shall use the subscript "$\infty$" to indicate the steady state distribution. 
Any other solution $x_n(t)$ goes to $X_n$ in the long time 
limit, while all the other modes are decaying.   
To find $X_n$ the equation should be solved 
with the boundary condition ${X_{\infty}=0}$, 
and normalized such that ${X_{0}=1}$. 
Clearly this is not required in practice:
because we already know the steady state solution \Eq{eq:fokker-planck-phi}, hence \Eq{eq:X-n-result}.

We define $x_n(t;\varphi_0)$ as the time-dependent solution 
for an initial preparation ${\rho_0(\varphi) = \delta(\varphi-\varphi_0)}$. 
Then we can express the correlation function of \Eq{e15}
as follows:
\beq
\label{eq:Cn-t-using-cond-prob}
C_n(t) \ = \ \braket{ x_{n}(t;\varphi) \cos(2\varphi) }_{\infty} - X_n X_1
\eeq
By linearity the $C_n(t)$ obey the same equation of motion 
as that of the $x_n(t)$, but with the special initial conditions ${C_n(0)=\Delta_n}$. 
Note that ${C_0(t)=0}$ at any time. In the infinite time limit ${C_n(\infty)=0}$ for any $n$.

Our interest is in the area $c_n$ as defined in \Eq{eq:cn-integral}.
Writing \Eq{eq:d-dt-xn-recursive} for $C_n(t)$, and integrating 
it over time we get 
\be{29}
\Lambda_n \, c_n \ - \ W_n  \left( c_{n-1} - c_{n+1} \right) \ \ = \ \ \Delta_n 
\eeq
This equation should be solved with the boundary conditions ${c_0=0}$ and ${c_{\infty}=0}$.
The solution is unique because the $n=0$ site has been effectively removed, 
and the truncated matrix is no longer with zero mode.  
One possible numerical procedure is to start iterating with $c_1$ as initial condition, 
and to adjust it such that the solution will go to zero at infinity. 
An optional procedure is to integrate the recursion backwards as explained 
in the next section. The bottom line is the following expression 
\beq
\label{eq:Dr-exact-result}
D_r \ \ = \ \ c_1 w^2 \ \ = \ \ - \sum_{n=1}^{\infty} \frac{(-1)^{n}}{n} \Delta_n X_n w
\eeq
where $X_n$ and $\Delta_n$ are given by \Eq{eq:X-n-result} and \Eq{eq:delta-n-def} respectively.

The leading term approximation ${D_r \approx  \Delta_1 X_1 w}$
is consistent with the heuristic expression ${D_r \approx (1/2) \Delta_1 \tau w^2}$ 
of \Eq{eq:dr-large-d-approximation} upon the identification 
\beq
\label{eq:tau-const-from-exact-result}
\tau \ \ = \ \ \frac{2}{w} \left[1-\exp\left(-\frac{w}{4D}\right)\right]
\eeq 
This expression reflects the crossover from diffusion-limited (${\tau \propto 1/D}$) 
to drift-limited  (${\tau \propto 1/w}$) spreading. 
\Fig{fig:dr-vs-w/d} compares the approximation that is based 
on \Eq{eq:dr-large-d-approximation} with \Eq{eq:tau-const-from-exact-result} 
to the exact result \Eq{eq:Dr-exact-result}. 

In the limit $(w/D)\rightarrow 0$ the asymptotic result for the radial 
diffusion coefficient is ${D_r=w^2/(8D)}$. We now turn to figure out 
what is the asymptotic result in the other extreme limit ${(w/D)\rightarrow\infty}$.  
The large $w/D$ approximation that is based on the first term of \Eq{eq:Dr-exact-result}, 
with the limiting value ${X_1=1}$, provides the asymptotic estimate ${D_r \approx  2D^2/w}$.
This expression is based on the asymptotic result \Eq{e19} for $\Delta_n$ with ${n=1}$.
In fact we can do better and add all the higher order terms. Using Abel summation we get  
\beq
D_r \ = \ 2\frac{D^2}{w} \sum_{n=1}^{\infty} (-1)^{n{-}1} n \ = \  \frac{1}{2}\frac{D^2}{w}
\eeq
Thus the higher order terms merely add a factor $1/4$ to the asymptotic result.
If we used \Eq{eq:dr-large-d-approximation}, 
we would have obtained the wrong prediction $D_r \approx D/2$
that ignores the $\tau$ dependence of \Eq{eq:tau-const-from-exact-result}.

\section{Derivation of the recursive solution}

In this section we provide the details of 
the derivation that leads from \Eq{e29} to \Eq{eq:Dr-exact-result}.
We define ${W_n^{\pm}= \mp W_n}$ and rewrite the equation 
in the more general form  
\beq
-W_n^{+}c_{n+1} + \Lambda_n c_n - W_{n}^{-}c_{n-1} \ = \ \Delta_n
\eeq
A similar problem was solved in \cite{Coffey1993},  
while here we present a much simpler treatment.
First we solve the associated homogeneous equation.
The solution $c_n=X_n$ satisfies   
\beq
-W_n^{+}X_{n+1} + \Lambda_n X_n - W_{n}^{-}X_{n-1} \ = \ 0
\eeq
and one can define the ratios ${R_n = X_n/X_{n-1}}$. 
Note that these ratios satisfies a simple  first-order recursive relation.  
However we bypass this stage because we can extract 
the solution from the steady state distribution. 

We write the solution of the non-homogeneous equation as 
\beq
c_n \ \ := \ \  X_n \tilde{c}_n
\eeq
and we get the equation 
\beq \nonumber
-W_n^{+} X_{n+1} \tilde{c}_{n+1} + \Lambda_n X_n \tilde{c}_n - W_{n}^{-}X_{n-1}\tilde{c}_{n-1} \ = \ \Delta_n
\eeq
Clearly it can be re-written as 
\beq  \nonumber
-W_n^{+} X_{n+1} (\tilde{c}_{n+1}-\tilde{c}_{n}) + W_{n}^{-}X_{n-1}(\tilde{c}_{n}-\tilde{c}_{n-1}) \ = \ \Delta_n
\eeq
We define the discrete derivative
\beq 
\tilde{a}_{n}  \ \ := \ \ \tilde{c}_{n}-\tilde{c}_{n-1}
\eeq
And obtain a reduction to a first-order equation: 
\beq
-W_n^{+} X_{n+1} \tilde{a}_{n+1} +  W_{n}^{-}X_{n-1} \tilde{a}_{n} \ = \ \Delta_n
\eeq 
This can be re-written in a simpler way by appropriate 
definition of scaled variables.   
Namely, we define the notations 
\beq
\tilde{R}_n \ = \ \frac{W_n^{+}}{W_n^{-}} R_n 
\hspace{2cm} 
\tilde{\Delta}_n \ = \ \frac{\Delta_n}{W_n^{+}}
\eeq 
and the rescaled variable 
\beq
a_n \ \ :=  \ \ X_n \tilde{a}_n
\eeq
and then solve the $a_n$ recursion in the backwards direction:  
\beq
a_{\infty}=0; 
\ \ \ \ a_n = \tilde{R}_n \left[\tilde{\Delta}_n +  a_{n+1} \right]
\eeq
If all the $R_n$ were unity it would imply that $a_1-a_{\infty}$ equals $\sum \Delta_n$.   
So it is important to verify that the "area" converges.
Next we can solve in the forward direction the $c_n$ recursion 
for the non-homogeneous equation, namely,  
\beq
c_0=0; \ \ \ \ \  c_n = R_nc_{n-1} + a_n 
\eeq
In fact we are only interested in 
\beq
c_1 \ \ = \ \ a_1  \ \ = \ \  \tilde{R}_1 \tilde{\Delta}_1 +  \tilde{R}_1  \tilde{R}_2 \tilde{\Delta}_2 + ... 
\eeq 
Note that in our calculation the $\tilde{R}_n=-R_n$, 
and therefore ${\tilde{R}_1\cdots\tilde{R}_n = (-1)^n X_n}$.

\section{The moments of the radial spreading}

\begin{figure}[t]
\includegraphics[width=7cm]{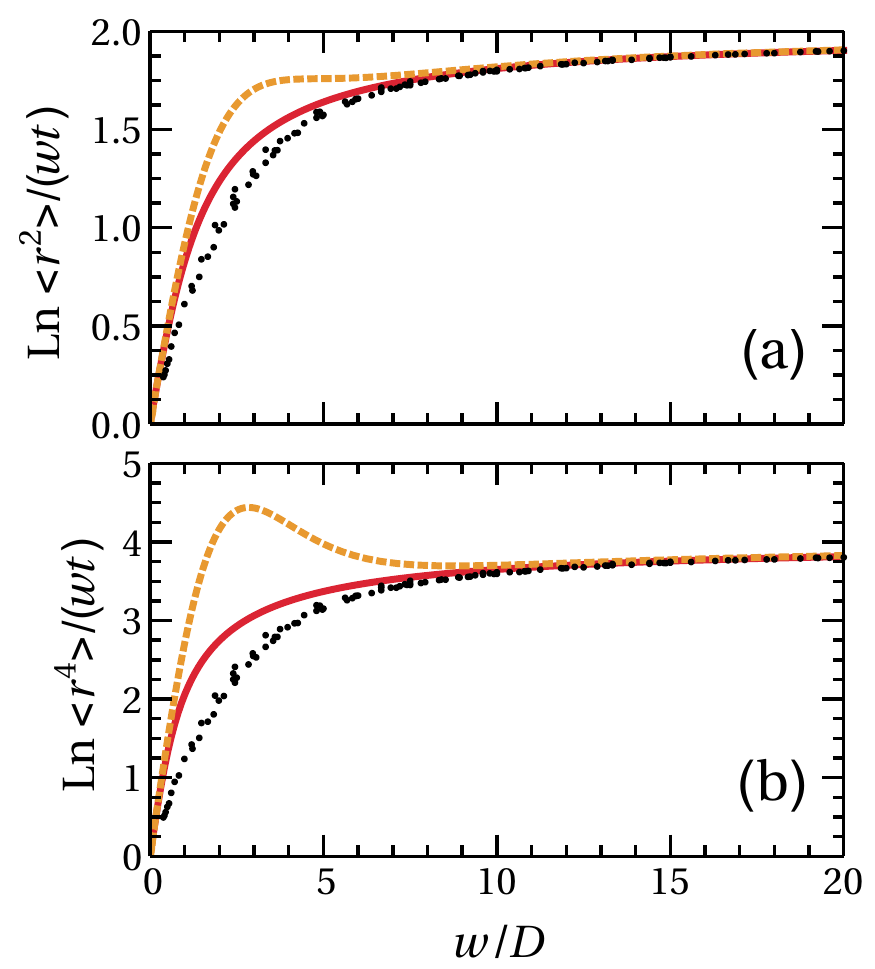}
\caption{ 
Scaled moments versus $w/D$.
The red solid lines are the exact results for the 2nd and 4th moments,
given by \Eq{eq:d-dt-ln-avg-r2-long-time} and \Eq{eq:d-dt-r4-matrix}, 
and the large $w/D$ asymptotic values are at $2$ and $4$, respectively.
These are compared with the numerical results (black symbols), 
and contrasted with the Lognormal prediction (orange dashed lines).
The later provides an overestimate for intermediate values of $w/D$.
} 

\label{fig:2-and-4-moments}
\end{figure}

The moments of a lognormal distribution are given by the following expression
\be{40}
\label{eq:log-normal-moments}
\ln \langle r^n \rangle \ \ = \ \ \mu n \ + \ \frac{1}{2}\sigma^2 n^2  
\eeq
On the basis of the discussion after \Eq{eq:langevin-ln-r}, 
if one assumed that the radial spreading at time~$t$ 
could be {\em globally} approximated by the lognormal distribution (tails included), 
it would follow that 
\be{41}
\frac{d}{dt} \ln \langle r^n \rangle \ \ = \ \ n w_r + n^2 D_r 
\eeq
In \Fig{fig:2-and-4-moments} we plot the lognormal-based expected growth-rate 
of the 2nd and the 4th moments as a function of $w/D$.
For small $w/D$ there is a good agreement with the expected results, 
which are $w^2/D$ and  $3w^2/D$ respectively. 
For large $w/D$ the dynamics is dominated by the stretching, 
meaning that $w_r \approx  w$, while ${D_r \rightarrow  0}$, 
so again we have a trivial agreement.
But for intermediate values of $w/D$ the lognormal moments 
constitute an overestimate when compared with the 
exact analytical results that we derive in the next section. 
In fact also the exact analytical result looks like an 
overestimate when compared with the results of numerical 
simulations. But the latter is clearly a sampling issue   
that is explained in Appendix~D.

\begin{figure}[t]
\includegraphics[width=7cm]{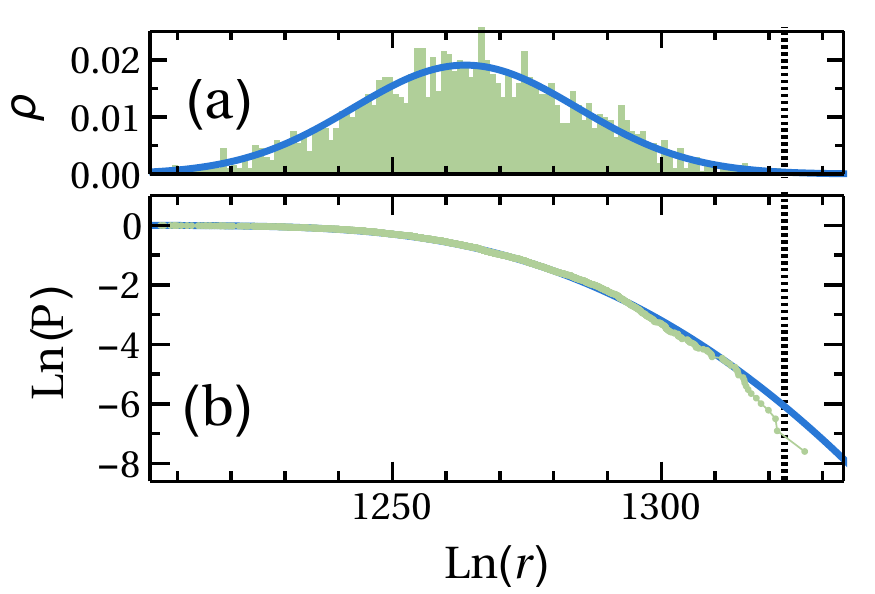}

\caption{ 
{\bf (a)} Distribution of $\ln(r)$ for $w/D=10/3$ after time $wt=2000$ 
with initial conditions $r=1$ and $\varphi = 0$. 
Numerical results (green histogram), that are based on $2000$ realizations,
are fitted to a Gaussian distribution (blue line). 
{\bf \ \ (b)}~Inverse cumulative probability of the same distribution.
The black dotted line indicate the numerically 
determined value ${\ln \braket{r^2}^{1/2} \approx  1323}$.
This value is predominated by the tail of the distribution.
The Gaussian fit fails to reproduce this value, 
and provides a gross over-estimate ${\ln \braket{r^2}^{1/2} \approx  1701}$.
\hfill
} 

\label{fg-distribution} 
\end{figure}

The deviation of the lognormal moments from the exact results 
indicates that the statistics of large deviations 
is not captured by the central limit theorem.
This point is illuminated in \Fig{fg-distribution}. 
The Gaussian approximation constitutes a good approximation 
for the body of the distribution but not for the tails 
that dominate the moment-calculation. 
Clearly, the actual distribution can be described as 
a {\em bounded} lognormal distribution, 
meaning that it has a natural cutoff which is implied by the strict inequality ${w_r<w}$. 
The stretching rate cannot be faster than~$w$.    
But in fact, as observed in \Fig{fg-distribution}b, 
the deviation from the lognormal distribution happens 
even before the cutoff is reached.

Below we carry out an exact calculation for the 2nd and 4th moments. 
In the former case we show that 
\be{42}
\label{eq:d-dt-ln-avg-r2-long-time}
\frac{d}{dt} \ln \langle r^2 \rangle  \ \ \sim \ \  2\left((w^2+D^2)^{1/2}-D\right) 
\eeq
This agrees with the lognormal-based prediction $w^2/D$ for ${(w/D)\ll 1}$, 
and goes to $2w$ for ${(w/D)\gg 1}$, as could be anticipated.

Before we go the derivation of this result we would like to illuminates 
its main features by considering a simple-minded reasoning. Let us ask ourselves 
what would be the result if the spreading was isotropic ($w_r = 0$). 
In such case the moments of spreading can be calculated as if 
we are dealing with the  multiplication of random numbers.
Namely, assuming that the duration of each step is ${\tau=1/(2D)}$, 
and treating~$t$ as a discrete index, \Eq{eq:langevin-r} implies that the spreading 
is obtained by multiplication of uncorrelated stretching factors ${\exp[w\tau\cos(\varphi)]}$.
Each stretching exponent has zero mean and dispersion ${\sigma_1^2=(1/2)[w\tau]^2}$, 
which implies $D_r = \sigma_1^2/(2\tau)$. Consequently we get 
for the moments    
\beq 
\langle r^n \rangle \ \ = \ \ 
\left[\braket{\eexp{nw\tau\cos(2\varphi)} } \right]^{t/\tau} \ r_0^n
\eeq
leading to 
\be{44}
\frac{d}{dt} \ln \langle r^n \rangle  \ \ = \ \
\dfrac{1}{\tau} \ln\left[I_0 \left(\sqrt{2} n \sigma_1 \right)\right]  
\eeq
This gives a crossover from $n^2D_r$ for ${\sigma_1\ll 1}$ 
to $nw$ for ${\sigma_1\gg 1}$, reflecting isotropic 
lognormal spreading in the former case, and pure stretching
in the latter case. So again we see that the asymptotic 
limits are easily understood, but for the derivation of 
the correct interpolation, say \Eq{eq:d-dt-ln-avg-r2-long-time}, 
further effort is required.

\section{The exact calculation of the moments}

We turn to perform an exact calculation of the moments. 
One can associate with the Langevin equation \Eq{e1} 
an FPE for the distribution, and from that to derive 
the equation of motion for the moments.
The procedure is explained and summarized in Appendix~E.
For the first moments we get 
\beq
\frac{d}{dt}\braket{x} &=& w \braket{x} - D \braket{x}\\
\frac{d}{dt}\braket{y} &=& - w \braket{y} - D \braket{y}
\eeq
with the solution
\beq
\braket{x} &=&   x_0 \,\exp [-(D-w) t] \\
\braket{y} &=&   y_0 \,\exp [-(D+w)t]
\eeq
For the second moments 
\beq
\frac{d}{dt}\left(\begin{array}{c}
\braket{x^2}\\
\braket{y^2}
\end{array}\right) 
&=& 
\Big[-2 D + 2D \bm{\sigma}_1  + 2 w \bm{\sigma}_3\Big]
\left(\begin{array}{c}
\braket{x^2}\\
\braket{y^2}
\end{array}\right)
\ \ \ \ 
\\
\frac{d}{dt}\braket{x y} \ \ 
&=& -4 D \braket{x y}
\eeq
where $\bm{\sigma}$ are Pauli matrices.
The solution is:
\beq
\left(\begin{array}{c}
\braket{x^2}\\
\braket{y^2} \\
\braket{xy}
\end{array}\right) 
=
\left[
\amatrix{e^{-2 D t} \bm{M} & 0 \cr 0 & e^{-4 D t}} 
\right]
\left(\begin{array}{c}
x_0^2\\
y_0^2 \\
x_0 y_0
\end{array}\right)
\eeq
where $\bm{M}$ is the following matrix: 
\beq \nonumber
\cosh[2 (w^2 {+} D^2)^{1/2} t] 
+ \sinh[2 (w^2 {+} D^2)^{1/2} t] 
\frac{D \bm{\sigma}_1 + w \bm{\sigma}_3}{\sqrt{w^2 {+} D^2}}
\eeq
For an initial isotropic distribution 
we get ${\langle r^2 \rangle_t =  M r_0^2}$, 
where
\be{58} 
M &=&  
e^{-2 D t} \cosh [2 (w^2 + D^2)^{1/2} t]  
\\ \nonumber
&& + \frac{D}{\sqrt{w^2 + D^2}}
e^{-2 D t} \sinh{[2 (w^2 + D^2)^{1/2}}t]
\eeq
The short time $t$ dependence is quadratic, reflecting ``ballistic" spreading, 
while for long times
\beq
\nonumber
\braket{r^2}_t  &\approx&   
\frac{r_0^2}{2}
\ \left(1 + \frac{D}{\sqrt{w^2 {+} D^2}}\right) \times
\\ 
&& \ \exp\left[  2\left((w^2+D^2)^{1/2}-D\right) t \right] \label{eq:r2-exact}
\ \ \ \ 
\eeq
From here we get \Eq{eq:d-dt-ln-avg-r2-long-time}. 
For the 4th moments the equations are separated 
into two blocks of even-even powers and odd-odd powers in $x$ and $y$. 
For the even block:
\beq
\label{eq:d-dt-r4-matrix}
&&\frac{d}{dt}\begin{pmatrix}
\braket{x^4}\\
\braket{x^2 y^2}\\
\braket{y^4}\\
\end{pmatrix} 
\ = \ 
2\tilde{\bm{M}} 
\begin{pmatrix}
\braket{x^4}\\
\braket{x^2 y^2}\\
\braket{y^4}\\
\end{pmatrix}
\eeq
where
\beq
\tilde{\bm{M}} 
\ = \ 
\begin{pmatrix}
 2 (w{-}D) & 6D & 0 \\
 D & -6D & D \\
 0 & 6D & -2(w{+}D) \\
\end{pmatrix}
\eeq
The eigenvalues of this matrix are the solution 
of $ \lambda^3 + 10 D \lambda^2 + (16 D^2 - 4 w^2) \lambda - 24 D w^2  = 0$. 
There are two negative roots, and one positive root. 
For small $w/D$ the latter is ${\lambda \approx (3/2)(w^2/D)}$, 
and we get that the growth-rate is ${3w^2/D}$
as expected from the log-normal statistics.

\hide{
Note: The ``averaged'' process H: $\dot{\vec{\braket{X}}} = H \vec{\braket{X}}$ is invariant under flipping of either the $x$ or $y$ axes (two commuting operator). As a consequences it is block-diagonal in a basis that is that is diagonal in either operators. It means for example that $\braket{x^2 y^2}$ will not be connected through $H$ to $\braket{x y^3}$, or that $\braket{x^2 y^3}$ will not be connected to $\braket{x^3 y^2}$.
}

\section{Discussion}

In this work we have studied the statistics of a stochastic squeeze process, 
defined by \Eq{e1}. Consequently we are able to provide a quantitatively valid
theory for the description of the noise-affected decoherence process 
in bimodal Bose-Einstein condensates, aka QZE. 
As the ratio $w/D$ is increased, the radial diffusion coefficient 
of $\ln(r)$ changes in a non-monotonic  
way from ${D_r=w^2/(8D)}$ to ${D_r=D^2/(2w)}$, 
and the non-isotropy is enhanced, namely 
the average stretching rate increases  
from ${w_r=w^2/(4D)}$ to the bare value ${w_r=w}$. 
The analytical results  \Eq{e16} and \Eq{eq:Dr-exact-result}
are illustrated in \Fig{fig:wr-vs-d} and \Fig{fig:dr-vs-w/d}, 

Additionally we have solved for the moments of~$r$. 
One observes that the central limit theorem is not enough 
for this calculation, because the moments are predominated 
by the non-Gaussian tails of the $\ln(r)$ distribution.
In particular we have derived for the second moment 
the expression ${\langle r^2 \rangle_t =  M r_0^2}$ 
with $M$ that is given by \Eq{e58}, or optionally one can use 
the practical approximation \Eq{eq:d-dt-ln-avg-r2-long-time}.


The main motivation for our work comes form the interest in the BJJ.
Form mathematical point of view the BJJ can be regarded 
as a quantum pendulum. It has both stable and unstable fixed points.
Its dynamics has been explored by numerous experiments. We mention 
for example Ref.\cite{Oberthaler2005} who observed both Josephson oscillations 
(``liberations") and self trapping (``rotations"), and Ref.\cite{Steinhauer2007}
who observed the a.c. and the d.c. Josephson effects. 
The phase-space of the device is spherical, known as the Bloch sphere.
A quantum state corresponds to a quasi-distribution (Wigner function) on that sphere, 
and can be characterized by the Bloch vector~$\vec{S}$. 
The length $\mathcal{F}=\left|\vec{S}\right|$ of the Bloch vector 
reflects the one-body coherence, and has to do with the ``fringe visibility"  
in a ``time-of-flight" measurement.  If all the particles are initially condensed 
in the upper orbital of the BJJ, it corresponds to a coherent ${\mathcal{F}=1}$ 
wavepacket that is positioned on top of the hyperbolic point, 
which corresponds to the upper position of the pendulum.
The dynamics has been thoroughly analyzed in \cite{Chuchem10} 
and experimentally demonstrated in \cite{chapman}.

To the best of our knowledge neither the Kapitza effect \cite{kapitza} 
nor the Zeno effect have been demonstrated experimentally in the BJJ context.
We expect the decay of~$\mathcal{F}$ to be suppressed  
due to the periodic or the noisy driving, respectively.
Let us clarify the experimental significance 
of our results for the full statistics of the radial spreading
in the latter case.  
In order to simplify the discussion, let us assume 
that the definition of $\mathcal{F}$ is associated 
with the measurement of a single coordinate~$\hat{x}$.
Measurement of~$\hat{x}$ is essentially the same 
as probing an occupation difference.
In a semiclassical perspective (Wigner function picture) 
the phase-space coordinate $x$ satisfies \Eq{e1}, 
where $\omega(t)$ arises from frequent interventions, 
or measurements, or noise that comes from the surrounding.  
Using a Feynman-Vernon perspective, each~$x$ outcome 
of the experiment can be regarded as 
the result of one realization of the stochastic process. 
The ``coherence" is determined by the second moment of $\hat{x}$. 
But it is implied by our discussion of the {\em sampling problem}  
that it is impractical to determine this second moment 
from any realistic experiment (rare events are not properly accounted).
The reliable experimental procedure would be to keep 
the {\em full} probability distribution of the measured $x$~variable, 
and to extract the $\mu$ and the $\sigma$ that characterize  
its lognormal statistics. For the latter we predict non-trivial dependence on~$w/D$.

Still, from purely mathematical point of view, 
one might be curious about the validity of the 
heuristic QZE expression \Eq{e3}. We already 
pointed out in the Introduction that 
the lognormal assumption implies that 
it should be replaced by \Eq{eLN}, which 
reduce to \Eq{e3} only for short times if 
the noise is very strong (small $w/D$).
We note that the expression that has been advertised 
originally in \cite{KhripkovVardiCohen2012} was 
slightly different, namely,   
\be{LNo}
\mathcal{S}(t) \ = \  e^{4D_r t} \cosh(2w_r t) - 1  
\eeq
The difference is due to the assumption (there) that 
it is $\alpha$, as defined in Appendix~A,  
rather than $r$ that has a lognormal distribution.
In physical terms it is like ignoring the initial 
isotropy of the preparation, hence creating 
an artifact - an artificial transient.  
In any case we found in the present work that 
none of these expressions are correct. 
This is because the tail of the distribution is bounded.
From \Eq{eq:d-dt-ln-avg-r2-long-time} we deduce that 
a practical approximation would be 
\be{LNn}
\mathcal{S}(t) \ = \ 2\left((w^2+D^2)^{1/2}-D\right) \, t
\eeq
Note that both expression \Eq{eLN} and \Eq{eLNn}
agrees with the heuristic expectation $(w^2/D) t$ for ${(w/D)\ll 1}$, 
and goes to bare non-suppressed value $2wt$ for ${(w/D)\gg 1}$. 
The difference between them is for intermediate values 
of $w/D$ where the lognormal prediction is an overestimate.
On the other hand, in a realistic experiment, we expect 
an underestimate as illustrated in \Fig{fig:2-and-4-moments}.

\appendix 

\section{The squeeze operation}

The squeeze operation is described by a real symplectic matrix 
that has unit determinant and trace ${|a|>2}$. Any such matrix 
can be expressed as follows: 
\beq
\bm{U} \ = \ \left(\amatrix{a & b \\ c & d}\right)  \ = \ \pm \eexp{\alpha \bm{H}} 
\ \ \ \ \ \ \ \ [ad-cb=1]
\eeq
where $\bf{H}$ is a real traceless matrix that satisfies ${\bm{H}^2=1}$.  
Hence it can be expressed as a linear combination of the three Pauli matrices: 
\beq
\bm{H} \ = \ n_1\bm{\sigma}_1 + i n_2\bm{\sigma}_2 +  n_3\bm{\sigma}_3  
\ \ \ \ 
\eeq
with ${n_1^2-n_2^2+n_3^2=1}$. Consequently 
\beq
\bm{U} \ \ = \ \ \pm \left[ \cosh(\alpha)\bm{1} + \sinh(\alpha) \bm{H} \right]
\eeq 
We define the canonical form of the squeeze operation as 
\beq
\bm{\Lambda} \ \ = \ \ \left(\amatrix{\exp(\alpha) & 0 \\ 0 & \exp(-\alpha)} \right) 
\eeq
Then we can obtain any general squeeze operation via similarity transformation 
that involves re-scaling of the axes and rotation, and on top an optional reflection.

We can operate with $\bm{U}$ on an initial isotropic cloud  
that has radius ${r_0=1}$. Then we get a stretched cloud 
that has spread ${\braket{r^2}=\mathcal{A} \, r_0^2}$, where 
\be{A5}
\mathcal{A} \ \ \equiv \ \  \left. \braket{r^2} \right|_{r_0{=}1} \ \ = \ \ \cosh(2\alpha)
\eeq
We also define the ``spreading" as  
\beq
\mathcal{S} \ \ = \ \ \mathcal{A}-1 \ \ = \ \ 2\sinh^2(\alpha)
\eeq
The notation $\alpha$ has no meaning for a {\em stochastic} squeeze process, 
while the notation ${\mathcal{A}\equiv \braket{r^2}}$ still can be used. 
In the latter case the average is over the initial conditions 
and also over realizations of~$\omega(t)$, 
implying that in \Eq{eA5} the~$\cosh(2\alpha)$ should be averaged over~$\alpha$.

\section{Numerical simulations}

\begin{figure}
\includegraphics[width=\hsize]{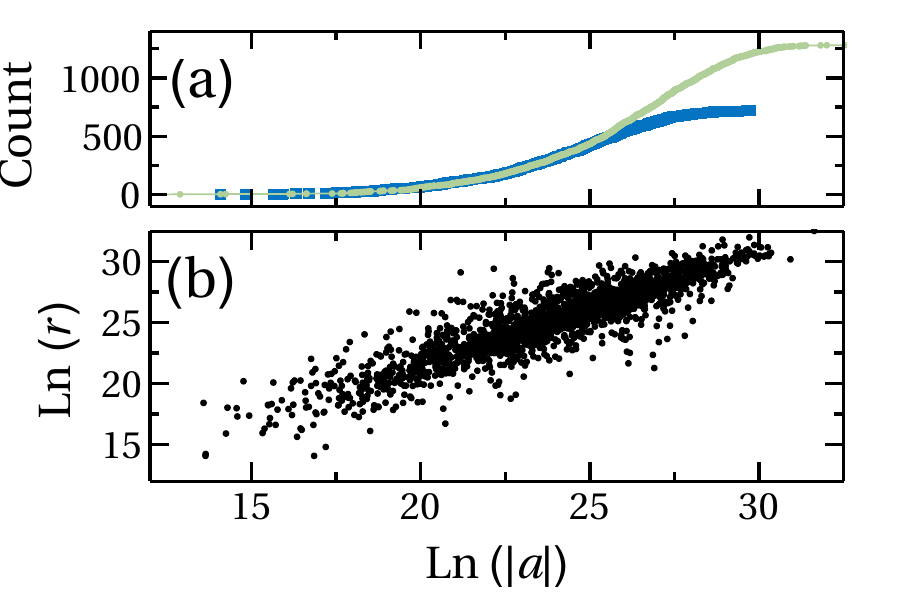}

\caption{ 
We consider 2000 realizations of a stochastic squeeze process. 
For each realization the trace ${a=\trc(\bm{U})}$ is calculated. 
{\bf (a)} The cumulative count of the $a$ values.  
Green points are for positive values,
while blue rectangles are for negative values. 
Here ${(w/D) = 10/3}$ and ${w t = 40}$. 
For simulations with longer times the distribution of 
positive and negative values become identical (not shown).
{\bf (b)} Scatter plot of $|a|$ versus the radial coordinate $r$.
For simulations with longer times we get full correlation.
\hfill
} 

\label{trace-U-comulative} 
\end{figure}

There are numerous numerical schemes that allow 
the simulation of a Langevin Equation. For example, 
the Milstein, the Runge-Kutta, and higher-order approximations 
such as the truncated Taylor expansion \cite{KP}.
These schemes are based on iterative integration of the Langevin equation,
then Taylor expand the solution in small~$dt$.
The dynamics generated by \Eq{e1} is symplectic,
however the numerical methods listed above do not respect this constraint.
Instead one can exploit the linear nature of the problem. 
Namely, \Eq{e1} is re-written as
\begin{align}\label{eq:lan-dynamic}
& \dot{\bm{r}}_t = \bm{H}(t) \bm{r}_t\\
& \bm{H} =  \bm{H}_s + \bm{H}_r(t)
\end{align}
Where $\bm{H}_s$ and $\bm{H}_r$ are the generators  
of the stretching and the angular diffusion, respectively,   
while $\bm{r}_t = (x_t, y_t)$.
If $\bm{H}_r$ were constant,
the solution of \Eq{eq:lan-dynamic} would be obtained 
by simple exponentiation of $\bm{H}$, 
namely $\bm{r}_{t_f} = \bm{U} \bm{r}_{0}$, 
with $ \bm{U} = \exp[(\bm{H}_r + \bm{H}_s)t_f]$. 
Choosing a small enough time interval $dt$ and
using the Suzuki-Trotter formula, the latter
equation is approximated by
\begin{align}
& \bm{U} = \bm{U}_{t_f} \cdots \bm{U}_{3dt} \bm{U}_{2dt} \bm{U}_{dt} \label{eq:u=ut}\\
& \bm{U}_t =  \exp{(\bm{H}_s dt)} \exp {(\bm{H}_r dt)} \label{eB6}
\end{align}
Where $\bm{U}_t$ gives the evolution of the vector $\bm{r}_t$ for
small time~$dt$, namely, ${\bm{r}_{t}=\bm{U}_t \bm{r}_{t-dt}}$.
\Eq{eq:u=ut} is valid also for time dependent $\bm{H}$, 
where the small step evolution \Eq{eB6} takes the form
\beq \label{eq:ut}
\bm{U}_t \ \ = \ \ 
\left(\begin{matrix}
e^{w \, dt} & 0\\
0 & e^{-w \, dt} 
  \end{matrix}\right)
\  
\left(\begin{matrix}
    \cos{\alpha_t} & -\sin{\alpha_t} \\
    \sin{\alpha_t} & \cos{\alpha_t} 
  \end{matrix}\right) 
\eeq
The uncorrelated random variables $\alpha_t$ have zero mean, 
and are taken from a box distribution of width $\sqrt{24 D \, dt}$, 
such that their variance is ${2D\, dt}$.   
As a side note we remark that by Taylor expanding \Eq{eq:ut} to
second order in $dt$, the Milstein scheme is recovered.
The radial coordinate $r$ is calculated 
under the assumption that the the preparation 
is ${(x_0{=}1,y_0{=}0)}$. Accordingly, 
what we calculate for each realization is 
\be{B3}
r \ \ = \ \ \sqrt{ U_{xx}^2 + U_{yx}^2 }  
\eeq      
In \Fig{trace-U-comulative}a we display the distribution of the trace $a$ for 
many realizations of such stochastic squeeze process.
Rarely the result is a rotation, and therefore in the main text 
we refer to it as ``squeeze".  
From the trace we get the squeeze exponent $\alpha$, 
and from \Eq{eB3} we get the radial coordinate~$r$. 
The correlation between these two squeeze measures 
is illustrated in \Fig{trace-U-comulative}b.
For the long time simulations that we perform in order to 
extract various moments, we observe full correlation (not shown).
In order to extract the various moments, we perform the simulation for a maximum time of $wt = 7500$,
with the initial condition $\bm{r}_0 = (1, 0)$.

We note that the results of Section~IX for the evolution of the moments 
can be recovered by averaging over product of the evolution matrices. 
For the first moments  we get the linear relation ${\braket{\bm{r}_t} = \braket{\bm{U}} \bm{r}_0}$, where   
\beq \nonumber 
\braket{\bm{U}} \ \ &=& \ \ \braket{ ... \ \bm{U}_{t_3} \ \bm{U}_{t_2} \ \bm{U}_{t_1} } 
\ \ = \ \ \left[ \braket{\bm{U}_t} \right]^{t/dt} 
\\
&& \ \ = \ \ 
\left(\begin{matrix}
e^{-(D+w)t} & 0\\
0 & e^{-(D-w)t} 
\end{matrix}\right) 
\eeq
Similar procedure can be applied for the calculation of the higher moments.

\section{Relation to QZE}

It is common to represent the quantum state of the  bosonic Josephson junction by a Wigner function on the Bloch sphere, see \cite{Chuchem10} for details. A coherent state is represented by a Gaussian-like distribution, namely
\be{328}
\rho^{(0)}(x,y) \ \ \approx \ \ 2\exp\left[-\frac{1}{\hbar}(x^2+y^2)\right]
\eeq 
where $x$ and $y$ are local conjugate coordinates.
The Wigner function is properly normalized with integration measure ${dxdy/(2\pi\hbar)}$.  
The dimensionless Plank constant is related to the number $N$ of Bonsons, namely ${\hbar=(N/2)^{-1}}$. 
After a squeeze operation one obtains a new state ${ \rho^{(t)}(x,y) }$.
The survival probability is 
\be{C2}
\mathcal{P}(t) = 
\tr\left[\rho^{(0)} \rho^{(t)}\right]
 = \frac{1}{\cosh(\alpha)} 
 = \frac{1}{1{+}\frac{1}{2}\mathcal{S}(t)} \ \ 
\eeq
However it is more common, both theoretically and experimentally to quantify the 
decay of the initial state via the length of the Bloch vector, 
namely ${\mathcal{F}(t)=|\vec{S}(t)|}$. 
It has been explained in \cite{KhripkovVardiCohen2012} that 
\be{C3}
\mathcal{F}(t) \approx  \exp\left\{- \hbar \sinh^2(\alpha) \right\}
= \exp\left\{- \frac{\hbar}{2}\mathcal{S}(t) \right\} \ \ 
\eeq
Comparing with the short time approximation of \Eq{eC2}, 
namely ${\mathcal{P}\approx \exp[-(1/2)\mathcal{S}(t)] }$, 
note the additional $\hbar=2/N$ factor in \Eq{eC3}. 
This should be expected: the survival probability drops to zero 
even if a single particle leaves the condensate. Contrary to that,  
the fringe visibility reflects the expectation value of the 
condensate occupation, and hence its decay is much slower. 
Still both depend on the spreading ${\mathcal{S}(t)}$.       

The dynamics that is generated by \Eq{e1} does not change 
the direction of the Bloch vector, but rather shortens its length, 
meaning that the one-body coherence is diminished, 
reflecting the decay of the initial preparation.
Using the same coordinates as in \cite{KhripkovVardiCohen2012} 
the Bloch vector is ${\vec{S}(t)=(S,0,0)}$, hence all the information 
is contained in the measurement of a single observable, 
aka fringe visibility measurement.

For a noiseless canonical squeeze operation 
we have ${D=0}$ and ${\alpha=wt}$, hence 
one obtains  ${\mathcal{S}(t)=2\sinh^2(w t)}$ 
which is quadratic for short times. 
In contrast to that, for a stochastic squeeze process \Eq{eC3}
should be averaged over realizations of $\omega(t)$. 
Thus $\mathcal{F}(t)$ is determined by the 
full statistics that we have studied in this paper.

At this point we would like to remind the reader what is the 
common QZE argument that leads to the estimate of \Eq{e3}. 
One assumes that for strong ${D}$ 
the time for  phase randomization is ${\tau =1/(2D)}$.
Dividing the evolution into $\tau$-steps,  
and assuming that at the end of each step 
the phase is totally randomized (as in projective measurement) 
one obtains
\beq
\overline{\mathcal{A}(t)} &\approx& \left[\overline{\mathcal{A}(\tau)}\right]^{t/\tau}
\ \ \approx \ \ \left[1-2(w\tau)^2\right]^{t/\tau} \\ 
\ \ &\approx& \ \ \exp\left[-(w^2/D)t\right]
\eeq 
The overline indicates average over realizations, as discussed after \Eq{eA5}.  
The short time expansion of exponent is linear rather than quadratic, 
and the standard QZE expression \Eq{e3} is recovered. 
This approximation is justified in the ``Fermi Golden rule regime", 
namely for ${\tau \ll  t  \ll  t_r}$, 
during which the deviation from isotropy 
can be treated as a first-order perturbation.
For longer times, and definitely for weaker noise, 
the standard QZE approximation cannot be trusted.

\section{Sample moments of a lognormal distribution}

\begin{figure}
\includegraphics[width=7cm]{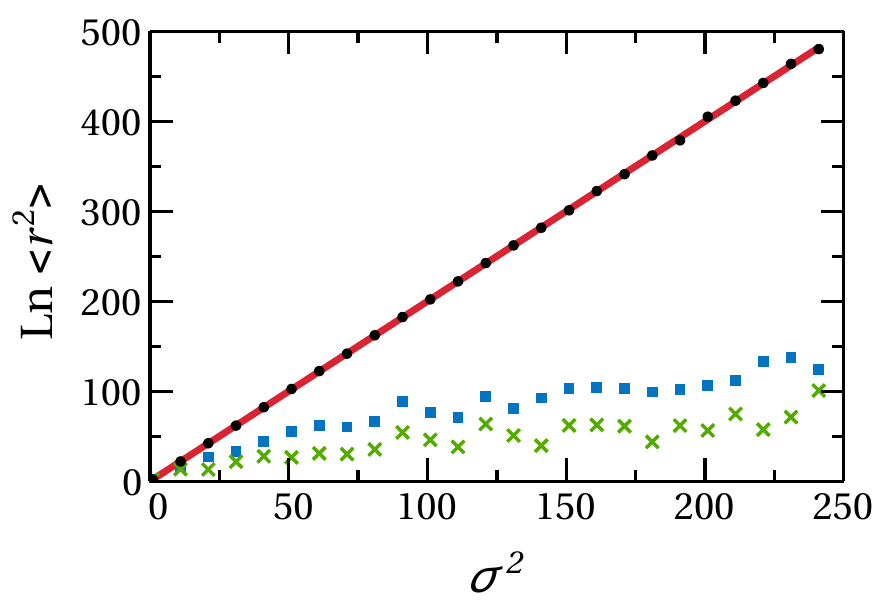}

\caption{ 
$\ln \braket{r^2}$ versus $\sigma$ for Lognormal distribution. 
Without loss of generality $\mu = 0$. 
The true result is represented by red line.
Numerical estimate based on $10^2$ and $10^5$ realizations
are indicated by green crosses and blue rectangles, respectively.
For the latter set of realization we get a much better estimate 
using an optional procedure (black dots). Namely, 
we calculate the sample average and the sample variance
of the $\ln r$ values in order to determine $\mu$ and $\sigma$,  
and then use \Eq{eq:log-normal-moments} to estimate the moments.
\hfill
} 
\label{fig:sample-moment-lognormal} 
\end{figure}

Consider a lognormal distribution of $r$ values. 
This mean that the $\ln r$ values have a Gaussian distribution.  
For a finite sample of $N$ values, one can calculate 
the sample average and the sample variance
of the $\ln r$ values in order to get a {\em reliable} 
estimate for $\mu$ and $\sigma$, 
and then calculate the moments $\braket{r^n}$ via \Eq{eq:log-normal-moments}. 
But a direct calculation of these moments provides 
a gross under-estimate as illustrated in \Fig{fig:sample-moment-lognormal}. 
This is because the direct average is predominated 
by rare values that belong to the tail of the distribution. 

The lesson is that direct calculation of moments 
for log-wide distribution cannot be trusted. 
It can provide a lower bound to the true results, 
not an actual estimate.

\section{Fokker-Planck from Langevin equation}
\label{sec:langevin-fokker}

We provide a short derivation for the FPE 
that is associated with a given Langevin equation.
From this we obtains the equations of motion for observables. 
For sake of generality we write the Langevin equation as follows:
\beq
\label{eq:lang-eq-multivariate}
&&\dot{x_j} \ = \ v_j + g_j \, \omega(t) \ \equiv \ f_j  
\\ \label{eCORR}
&&\braket{\omega(t)\omega(t')} \ = \ 2D \delta_{\tau} (t-t')
\eeq
The $v_j$ and the $g_j$ are some functions of the $x_i$.
\Eq{e1} is obtained upon the identification ${x_j=(x,y)}$ 
and ${v_j=(wx,-wy)}$, and ${g_j=(-y,x)}$.
The ``noise'' has zero average, namely $\braket{\omega(t)} = 0$, 
and is characterized by a correlation time~$\tau$.  
Accordingly the $\delta_{\tau}(t-t')$ has a short but finite width, 
which is later taken to be zero.

For a particular realization of the noise, 
the continuity equation for the Liouville distribution $\rho(x)$ reads:
\beq
\label{eq:dro-dt-non-averaged}
\frac{\partial\rho}{\partial t}  \ \ = \ \ -\frac{\partial}{\partial x_j} \left( f_j \rho \right)
\eeq
We are interested in $\rho(x)$ averaged over many-realizations of the noise $\omega$.
In its current form \Eq{eq:dro-dt-non-averaged} cannot be averaged, 
because $\rho$ and $f$ are not independent variables.
To overcome this issue \Eq{eq:dro-dt-non-averaged} is integrated iteratively.
To second order one obtains 
\beq
\label{eq:fp-rho-t-plus-dt-minus-rho-t}
&&\rho(t+dt) - \rho(t) \ = \ 
\\ \nonumber
&& -\int_t^{t{+}dt}  \!\!\!\!\! dt' 
\dfrac{\partial}{\partial x_j} f_j(t') 
\left[ \rho(t) - \int_{t}^{t'} \!\! dt^{\prime\prime} 
\dfrac{\partial}{\partial x_k} f_k(t^{\prime\prime})  \rho(t) \right] 
\eeq
Performing the average over realizations of the noise, 
non-vanishing noise-related term arise from 
the correlator of \Eq{eCORR}. Then performing the $dt^{\prime\prime}$ 
integral over the broadened delta one obtains a~$1/2$ factor. 
Dividing both sides by $dt$, and taking the limit ${dt \rightarrow \tau \rightarrow 0}$,
one obtains:
\begin{align}\label{eq:fp-stratonovich}
\pd{\rho}{t} = 
-\frac{\partial}{\partial x_j} 
\left[   v_j \rho  - g_j D \frac{\partial}{\partial x_i} \left(g_i \rho\right) \right]
\end{align}
Terms that originate from higher order iterations or moments  
are $\mathcal{O}(dt)$ or vanish in the $\tau\rightarrow0$ limit.    
\Eq{eq:fp-stratonovich} is the FPE that is associated with the
Stratonovich interpretation of \Eq{eq:lang-eq-multivariate}, 
see Eq(4.3.45) in p.100 of \cite{Gardiner1985b}.

An observable $X$ is a function of the $x$ variables. 
In order to obtain an equation of motion for $\braket{X}$, 
we multiply both sides of \Eq{eq:fp-stratonovich} by~$X$, 
and integrate over~$x$.
Using integration by parts, 
and dropping the boundary terms,  
we get the desired equation:
\beq 
\nonumber
\frac{d}{dt} \braket{X} \ \ &=& \ \ 
\left\langle   
\frac{\partial X}{\partial x_{j}} 
\left( v_j + \frac{\partial g_j}{\partial x_i} D g_i \right) \right\rangle  
\\  \label{eq:average-moments-from-langevin}
&&\ + \ 
\left\langle \frac{\partial^{2} X}{\partial x_i \partial x_j} g_j D g_i \right\rangle 
\eeq
In the main text we use this equation for the moments
of the distribution (${x,y,x^2,xy,y^2,x^4,x^2y^2,y^4}$).

\textit{Remark concerning various interpretation of the Langevin equation.-- }
The Langevin equation defined by \Eq{eq:lang-eq-multivariate} and \Eq{eCORR}, with $\tau \rightarrow 0$,
can be written as an integral equation:
\beq \label{eq:langevin-integrated}
x_j(t) - x_j(0) = \int_0^t v_j dt' + \int_0^t g_j dW(t) 
\eeq
where
\beq
&& W(t) = \int_0^t \omega(t') dt' \\
&& dW(t) = W(t+dt) - W(t)
\eeq
The second integral in \Eq{eq:langevin-integrated}, 
is interpreted as a Riemann--Stieltjes like integral \cite{Hanggi1978}:
\beq \nonumber
\int_0^t g_j dW(t) = \lim_{N \rightarrow \infty } 
\sum_{n}^N g_j(\bar{x})  \left[ W(t_n) - W(t_{n-1}) \right] 
\eeq
where 
\beq
\bar{x} \ = \ \lambda x_i(t_{n-1}) + (1-\lambda) x_i(t_n) 
\eeq
with ${0\! <\!\lambda\! <\! 1}$, and ${0 = t_0 < .. < t_N = t}$.
Because of the singular nature of the stochastic process $W(t)$,
the final result of this integral depends on the chosen value of $\lambda$.
Each choice provides a different ``interpretation'' of the Langevin equation \cite{Sokolov2010}:
for $\lambda = 1$, the equation is interpreted as ``It\^{o}'';
for $\lambda = 1/2$ it is interpreted as  ``Stratonovich'';
and for $\lambda = 0$ it is interpreted as the ``H{\"a}nggi--Klimontovich''.
Each interpretation produces a different FPE. 
The Stratonovich interpretation leads to \Eq{eq:fp-stratonovich}, 
while for the other interpretations the RHS of \Eq{eq:fp-stratonovich} 
is replaced with:
\begin{align}\label{eq:ito-hanggi-fp}
&-\frac{\partial}{\partial x_j} 
\left[   v_j \rho  -  D \frac{\partial}{\partial x_i} \left(g_j g_i \rho\right) \right]
 &\text{(It\^{o})} \\ 
&-\frac{\partial}{\partial x_j} 
\left[   v_j \rho  - g_j g_i D \frac{\partial}{\partial x_i} \left( \rho\right) \right]
 &\text{(H{\"a}nggi)}
\end{align}
In the specific case of \Eq{e1} with $g = (-y,x)$,
we have $\partial_i g_i \rho = g_i \partial_i \rho$.
Consequently the same FPE is obtained 
for both the Stratonovich and the H{\"a}nggi interpretations. 
We note that turning off the squeeze in \Eq{e1} ($w = 0$),
and using either of these interpretations, 
the FPE becomes:
\begin{align}\label{eq:ro-only-noise}
 \pd{}{t} \rho(x,y,t) = D \left(x \pd{}{y} - y \pd{}{x} \right)^2 = D \dfrac{\partial^2}{\partial \varphi^2} \rho
\end{align}
Which is clearly the required equation. However if one uses the It\^o 
prescription, an additional term appears in the FPE, namely, 
$ -D \partial_x (x \rho) - D \partial_y (y \rho) $.


\end{document}